\documentclass[11pt]{article} \usepackage{hyperref} \pdfoutput=1
\begin{document}

\title{Optimal chaotic mixing by two-dimensional Stokes flows}
\author{Qizheng Yan \& David Saintillan \\ 
 \\
 \vspace{6pt} Department of Mechanical Science and Engineering, \\ 
University of Illinois at Urbana-Champaign, Urbana, IL 61801, USA}
\maketitle
\begin{abstract} 
Numerous mixing strategies in microfluidic devices rely on chaotic advection by time-dependent body forces. The question of determining the required forcing function to achieve optimal mixing at a given kinetic energy or power input remains however open. Using finite-horizon optimal control theory, we numerically calculated general optimal mixing flows in a two-dimensional periodic geometry as truncated sums of $N$ time-modulated Fourier modes that satisfy the Stokes equations. These flows were determined to minimize a multiscale mixing norm for a passive scalar at the final time, given a constraint of constant kinetic energy. In this fluid dynamics video, we show the evolution of the passive scalar field as it is advected by the optimal mixing flows, for different values of the number $N$ of Fourier modes in the flow fields. Very efficient mixing is obtained when $N$ is large, corresponding to a large number of length scales in the flow fields. We also present movies showing the dynamics of a patch of passive particles as they are transported by the flows, and much better spreading of the patch is obtained for large values of $N$.
 \end{abstract}
 
  \end{document}